\documentclass[10pt,twocolumn,aps,prb,amsmath,amssymb,showpacs]{revtex4-1}
\usepackage{graphicx}
\pdfoutput=1

\newcommand{\bra}[1]{ {\langle #1 \vert }}
\newcommand{\ket}[1]{ {\vert #1 \rangle }}
\newcommand{\ave}[1]{ {\langle #1 \rangle }}
\newcommand{\braket}[2]{ {\langle #1 \vert #2 \rangle} }
\newcommand{\vac}{0}

\newcommand{\ul}{\underline{l}}
\newcommand{\ur}{\underline{r}}
\newcommand{\uu}{\underline{u}}
\newcommand{\ud}{\underline{d}}

\begin{document}

\title{Fermionic Implementation of Projected Entangled Pair States Algorithm}

\author{Iztok Pi\v{z}orn}
\author{Frank Verstraete}
\affiliation{Faculty of Physics, University of Vienna, Boltzmanngasse 5, A-1090 Wien}

\date{April 22, 2010}

\pacs{02.70.-c, 71.10.Fd, 03.67.-a}

\begin{abstract}
We present and implement an efficient variational method to simulate two-dimensional finite size fermionic quantum systems by fermionic projected entangled pair states. The approach differs from the original one due to the fact that there is no need for an extra string-bond for contracting the tensor network.
The method is tested on a bi-linear fermionic model on a square lattice for sizes up to ten by ten where good relative accuracy is achieved. 
Qualitatively good results are also obtained for an interacting fermionic system.
\end{abstract}

\maketitle

\section{Introduction}
The theoretical study of quantum many-body systems presents one of the most challenging tasks of condensed matter physics, computational physics and quantum chemistry. 
Several approaches have been proposed to study quantum many-body systems e.g. quantum monte carlo (QMC), dynamical mean field theory (DMFT), density-matrix renormalization group (DMRG) \cite{whiteprl69,schollwoeck}/tensor network methods \cite{vidalprl93,verstraeteprl93}.
The latter are best suited to describe physical systems of local Hamiltonians at zero temperature in one spatial dimension for which it was shown \cite{faith,schuchprl100,hastings} that they can be well approximated by matrix product states (MPS). 

A generalization of MPS algorithms to two spatial dimensions was given by projected entangled pair states (PEPS) \cite{peps,valentin} algorithms where the quantum state is described in terms of entangled pairs on lattice bonds. Those states capture the entanglement structure needed to represent states that obey an area law \cite{eisertarealaw}, and there are strong arguments why every ground state of a gapped two-dimensional local Hamiltonian can be efficiently represented as a PEPS \cite{hastings,kraus}.

All these methods were originally constructed to simulate quantum \emph{spin} systems whereas practical problems in condensed matter physics and quantum chemistry are more often of fermionic nature. A notorious example is the Fermi-Hubbard model which is believed to be a good candidate for high-temperature superconductivity. For local Hamiltonians in one spatial dimension, the distinction between spins and fermions is irrelevant as any physical fermionic model can be transformed by Jordan-Wigner transformation to a spin model where the locality of interaction is preserved. This is not the case in two-dimensional systems where such transformation would in general convert local interactions to non-local strings operators. 
In the case of the ladders it is in principle possible to use linear DMRG methods if all the symmetries are exploited \cite{whitescalapino} but such an approach is clearly not scalable.

Around the same time, two independent approaches to simulate two-dimensional fermionic systems were proposed: 
a generalization of  the multiscale entanglement renormalization (MERA) \cite{mera} to fermionic systems \cite{corboz0} and the description of 
two-dimensional fermionic systems in terms of fermionic projected pair states (fPEPS) \cite{kraus}.
In the latter, PEPS were generalized to fermionic systems in a natural way by considering entangled fermionic pairs instead of entangled spin pairs.
 It was also shown that this is a good ansatz and is in principle able to parametrize ground states of gapped fermionic models. 
 However, in the case of the fPEPS the sign problem was not solved in a completely satisfactory way as there was still a need for increasing the bond dimension with a factor of two if one were to contract this fermionic PEPS using the standard procedure. 
The efficient contraction of fermionic MERA was reinterpreted in terms of fermionic swap and jump rules \cite{eisert1,corboz1} which allowed to generalize the sign-free contraction also to arbitrary fermionic tensor networks, as first reported in Ref. \cite{corboz1} and subsequently in Ref. \cite{eisert2}. In these papers, it was also sketched how this formalism can be used to contract general fPEPS.
The first fPEPS simulations albeit without the sign-free contraction rules, were performed in Refs.\cite{Zhou,Zhou2} under the name Graded PEPS, and very promising numerical results were reported.
Finally, the full sign-free fPEPS algorithm for infinite lattices was implemented, together with interesting numerical results on interacting fermions and the t-J model, in Ref. \cite{corboz2} where also an explicit scheme for contracting finite-size fPEPS was given.
A crucial element in all those approaches was the realization of a simple fermionic swapping rule which will also play an important role in this paper where we focus on construction of the finite-size fermionic PEPS algorithm.

The most obvious advantage of the finite lattice PEPS over the infinite PEPS (iPEPS) algorithm is that no assumption of translation invariance symmetry is required. The finite size PEPS algorithm is therefore well suited to simulate physical systems with an unknown translation invariance pattern or translation non-invariant (disordered or noisy) systems. The only input information for the finite PEPS algorithm is the Hamiltonian operator and, possibly,  the parity of the ground state. However, if the symmetries are known, they can be embedded naturally \cite{bauer}.

In this paper, we address the fermionic PEPS \cite{kraus} entirely in terms of fermions without introducing any additional bonds between lattice sites but rather embedding all fermionic signs locally. 
The crucial element in such description is a fermionic rule used to swap two fermionic tensor operators \cite{corboz1,eisert2,corboz2}.
This way the complexity of the method exactly translates to the conventional PEPS for quantum spin systems (strictly speaking, it is even more efficient due to the parity constraints). 
We are able to efficiently calculate expectation values of arbitrary operators for a given fPEPS state and efficiently optimize the ground state approximation for an arbitrary fermionic system on a rectangular lattice.  
We test the method on an integrable quadratic model on a square lattice and compare the ground state energy and the total particle number to the exact values.

\section{Fermionic Projected Entangled Pair States}
We start with rewriting the original fPEPS ansatz \cite{kraus} in an alternative way which will allow fermionic manipulations and consequently sign-free contraction of fermionic PEPS states. 
As given in \cite{kraus}, a quantum state of a fermionic system on a square lattice can be described in terms of fermionic entangled pair states as
\begin{equation}
    \ket{\Psi} = 
    W_{\underline{\alpha},\underline{\beta},\underline{\gamma},\underline{\delta} }
    \prod_{i,j} Q_{i,j} \prod_{i,j} H_{i,j} \prod_{i,j} V_{i,j}  \ket{\vac}
\label{eq:originalfpeps}
\end{equation}
where to each site $(i,j)$ four auxiliary fermions are associated: $\alpha_{i,j}$, $\beta_{i,j}$, $\gamma_{i,j}$ and $\delta_{i,j}$, connecting the site to the respective left, right, upper and lower neighbor. The entangled pairs on the horizontal and vertical bonds are created (up to the normalization) by operators
$H_{i,j} = 1 + \beta_{i,j}^\dagger\alpha_{i,j+1}^\dagger$ and 
$V_{i,j} = 1 + \delta_{i,j}^\dagger\gamma_{i+1,j}^\dagger$, respectively,
and the projection to the space of physical fermions is given by projectors 
\begin{equation}
Q_{i,j} = 
A_{l r u d k}^{[i,j]} 
c_{i,j}^{\dagger\, k} \alpha^l_{i,j} \beta^r_{i,j} \gamma^u_{i,j} \delta^d_{i,j}.
\label{eq:projP}
\end{equation}
For brevity, we omit the summation symbol where it is understood that the summation takes place over all indices that appear both in sub-script and superscript. 
The expression is traced over the space of virtual fermions which is formally designated by 
the operator
$W_{\underline{\alpha},\underline{\beta},\underline{\gamma},\underline{\delta} }$ which mimics
the vacuum of virtual particles in the subscript, e.g.
$W_{\underline{\alpha}} \equiv \prod_{\nu} \alpha_{\nu}\alpha_{\nu}^\dagger$ and
$W_{\underline{\alpha},\underline{\beta},\underline{\gamma},\underline{\delta} } \equiv%
W_{\underline{\alpha}} W_{\underline{\beta}} W_{\underline{\gamma}}W_{\underline{\delta}}$. 
Note that $W_{\underline{\alpha}} = W_{\underline{\alpha}}^\dagger$.
We will use a dash notation when referring to sequences $\underline{m} \equiv (m_1, m_2, \ldots)$ or tensors $A_{\underline{m}} \equiv A_{m_1, m_2,\ldots}$ with the rank given by the context.

A fundamental feature of fermionic systems due to the causality is that the system is always in a state with a well defined parity, $(\forall \nu)\enskip \bra{\Psi} c_{\nu} \ket{\Psi}=0$. When the ansatz~(\ref{eq:originalfpeps}) is used to describe the ground state of a fermionic system, one can therefore assume that the projection operators $Q_{i,j}$ are either parity preserving ($P_{i,j}=0$) or parity violating ($P_{i,j}=1$) and can be described by only half of the tensor elements of $\underline{A}^{[i,j]}$, i.e.
$(l+r+u+d+k)\,\textrm{mod}\, 2 \neq P_{i,j}  \Rightarrow A_{l r u d k}^{[i,j]} =0$. The practical consequence of such assumption is that the projection operators~(\ref{eq:projP}) either commute or anti-commute.

Let us consider a $m{\times}n$ lattice of fermions and choose the row-major order
in~(\ref{eq:originalfpeps}) by multiplying the projection operators $Q_{\nu}$ by the
entanglement creation operators 
$H_{\nu}$ and $V_{\nu}$ in~(\ref{eq:originalfpeps}).
This results in a  description  
in terms of two types of virtual fermions $\alpha_{\nu}$ and $\gamma_{\nu}$ on
horizontal and vertical bonds, respectively, 
\begin{equation}
    \ket{\Psi} = W_{\underline{\alpha},\underline{\gamma}} A_{m,n} \cdots A_{1,n}\cdots A_{1,1} \ket{0}
\end{equation}
with in general non-commuting operators
\begin{equation}
A_{i,j} = 
A_{l r u d k}^{[i,j]} 
c_{i,j}^{\dagger\, k} 
\alpha_{i,j}^l \alpha_{i,j+1}^{\dagger\, r} \gamma_{i,j}^u \gamma_{i+1,j}^{\dagger\, d} .
\label{eq:Aij}
\end{equation}
of the same parity as the corresponding $Q_{i,j}$, i.e. 
either parity preserving ($P_{i,j}=0$) or swapping ($P_{i,j}=1$).
Again, we use the operator $W_{\underline{\alpha},\underline{\gamma}}$ to mimic the
contraction over virtual particles.

\subsection{Expectation values}
A starting point in the computation with fermionic tensor product states is the 
calculation of expectation values of arbitrary operators. 
Due to the linearity it is sufficient to
calculate the expectation value of an arbitrary \emph{product} operator
\begin{equation}
O = O_{m,n}\cdots O_{1,n}\cdots O_{1,1}
\label{eq:O}
\end{equation}
where $O_{i,j}$ are single-site operators of a well defined parity $p_{i,j}$. 
Explicitly, each $O_{i,j}$ can be either 
parity preserving ($p_{i,j}=0$) in which case it can be written as 
$O_{i,j} = O_{0,0}^{[i,j]} c_{i,j} c_{i,j}^\dagger + O_{1,1}^{[i,j]} c_{i,j}^\dagger c_{i,j}$
for some coefficients $O_{0,0}^{[i,j]}$ and $O_{1,1}^{[i,j]}$, 
or parity swapping ($p_{i,j}=1$) such as 
$O_{i,j} = O_{0,1}^{[i,j]} c_{i,j} + O_{1,0}^{[i,j]} c_{i,j}^\dagger$ 
for some coefficients $O_{0,1}^{[i,j]}$ and $O_{1,0}^{[i,j]}$.
The expectation value is formally written as 
\begin{equation}
\bra{\Psi} O \ket{\Psi} = 
\bra{\vac}
{A'}_{1,1}^\dagger \cdots {A'}_{m,n}^\dagger O 
A_{m,n}\cdots A_{1,1}\ket{\vac}
\label{eq:obs1}
\end{equation}
where the conjugated state $\bra{\Psi}$ is described by a complementary set of virtual fermions
designated by a prime, i.e. 
\begin{equation}
{A'}_{i,j}^\dagger = 
A_{l,r,u,d,k}^{[i,j] *}
{\alpha'}_{i,j}^{\dagger\, l} {\gamma'}_{i,j}^{\dagger\, u}
{\gamma'}_{i+1,j}^d
{\alpha'}_{i,j+1}^r c_{i,j}^k .
\end{equation}

Exact contraction of such a tensor network, albeit possible, is inefficient due to the contraction
order specified in~(\ref{eq:obs1}). In order to contract the fermionic tensor network efficiently, 
one must be able to first contract over the physical modes and then contract the double layer in an approximate way\cite{peps}. In both steps one must be able to swap the contraction order between two tensor operators sharing a common contraction leg which, as we will show, is possible due to the parity constraints in fermionic tensor network.
The latter step is performed by merging rows together and representing the double row by a single row, such that the horizontal bond dimensions remain finite.  
%
%
In order to contract over the physical modes in the first step, the tensor network must be written in a way where both tensors corresponding to a specific site appear together, such as 
\( A_{i,j}^\dagger O_{i,j} A_{i,j} \) which are of a globally defined parity $p_{i,j}$ regardless of parity $P_{i,j}$ of $A_{i,j}$. 
 Indeed, substituting 
the result by effective operators ${\tilde A}_{\nu}$ and ${\tilde A}_{\mu}$ using a fermionic rule explained in the following, that is exactly what we are able to achieve.
Let us in the following present a rule which will allow  us to rewrite e.g.
${A'}_{i,j}^\dagger {A'}_{i,j+1}^\dagger = {\tilde A'}_{i,j+1} {\tilde A'}_{i,j}$ which is needed to reverse the contraction order in the conjugate layer.  
An equivalent rule was already used in Refs. \cite{corboz1,corboz2,eisert2}.

\begin{figure}
\centering
\includegraphics[width=0.8\columnwidth]{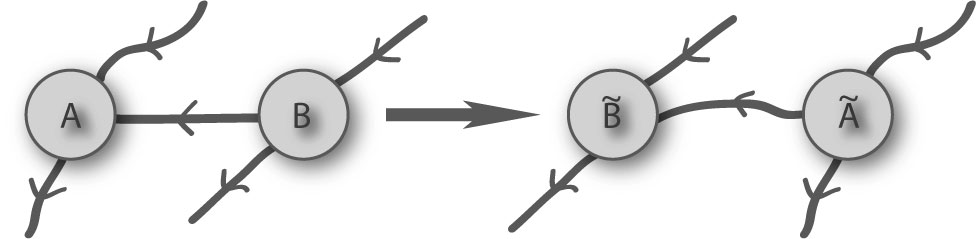}
\caption{Swapping of two operators $A B = {\tilde B} {\tilde A}$. }
\label{fig:swap}
\end{figure}
\emph{Fermionic swap rule}: let us define an arbitrary operator $A$ and 
an operator $B$ of well defined parity $p_B$ as
\begin{equation}
    A = A_{\ul\,\underline{a}\,\ur} \, 
    a_L^{\dagger\, \ul} \gamma^{\underline{a}} a_R^{\ur} %
    \quad\textrm{ and }\quad
    B = B_{\ul\,\underline{b}\,\ur} \, 
    b_L^{\dagger\, \ul} \gamma^{\underline{b}\, \dagger} b_R^{\ur}
\label{eq:fermionicrule1}
\end{equation}
where we use a notation 
\(  c^{\underline{m}} \equiv c_1^{m_1} c_2^{m_2} \cdots \) where $c_j$ represents fermionic annihilation
operators. Note that any superposition of products of fermionic operators can be written in this form. 
Then the following statement can be made:
t`he product $A B$ contracted over all common modes, 
here explicitly denoted by $\underline{\gamma} \equiv (\gamma_1, \gamma_2, \ldots)$,
can be written in a reverse order (\cite{eq:fermionicrule1}) as
\begin{equation}
W^\dagger A B W = W^\dagger  {\tilde B} {\tilde A} W
\quad\textrm{with} \quad 
W = W^\dagger = \prod_{j} \gamma_j \gamma_j^\dagger
\end{equation}
where ${\tilde A}$ and ${\tilde B}$ are obtained independently from
$A$ and $B$, respectively, in addition to a \emph{global} parity sign $p_B$, as 
\begin{eqnarray}
    {\tilde A} &=&  
\Big( A_{\ul\,\underline{a}\,\ur} (-1)^{(\underline{l} + \underline{a} + \underline{r}) p_B} \Big)
    \, a_L^{\ul} \gamma^{\underline{a}\, \dagger} a_R^{\ur} %
    \label{eq:swaprule} \\
{\tilde B} &=& 
    \Big( B_{\ul\,\underline{b}\,\ur}(-1)^{\underline{b}}\Big)
    \, b_L^{\ul} 
    \gamma^{\underline{b} }
    b_R^{\ur}.
\nonumber
\end{eqnarray}
where $(-1)^{\underline{x}} = (-1)^{x_1 + x_2 + \cdots}$.

The fermionic swap rule can be proven in a straight-forward way by writing the above ansatz and reordering the fermionic operators where all fermionic signs cancel except for the parity sign of the contraction product operator.
It is however crucial to assume that at least one of the operators is of well defined parity.
Otherwise, such decomposition is impossible since $(-1)^{x y}$ cannot be decomposed to  
a product $f(x) g(y)$ for $x, y \in \{0,1\}$ and any functions $f$ and $g$.

In the context of this paper, both $A$ and $B$ will always be of well defined parity. 
In such a case, the sign factor in ${\tilde A}$ in~(\ref{eq:swaprule}) becomes a globally defined quantity~$(-1)^{p_A p_B}$ which agrees with the sign produced by commuting $A$ and $B$ when they share no common fermionic mode. 

The only assumption in the above presented fermionic rule is that one of the operators is of a well defined parity. Therefore all fermionic swap rules in Refs.~\cite{corboz1,eisert2,corboz2}  with more severe constraints (either both $A$ and $B$ are of well defined parity or even both are parity preserving) necessarily coincide with the fermionic rule presented here.

Since the parity of $A_{i,j}$ (and thus ${A'}_{i,j}^\dagger$) is well defined by definition,
one may use the swap rule to reverse the contraction order of ${A'}_{i,j}^\dagger$ in 
(\ref{eq:obs1}) such that
\[
{A'}_{1,1}^\dagger \cdots {A'}_{m,n}^\dagger = 
(-1)^{f(P_{1,1}, \ldots, P_{m,n} ) }
{\tilde A'}_{m,n}\cdots {\tilde A'}_{1,1}
\]
where operators ${\tilde A'}_{i,j}$ are obtained from ${A'}^\dagger_{i,j}$ by absorbing the
local fermionic sign factor $(-1)^{l+u}$ arising from swapping two operators acting on a common virtual fermion,
\begin{equation}
    {\tilde A'}_{i,j} = A_{l r u d k}^{[i,j] *} (-1)^{l+u}
    {\alpha'}_{i,j}^{l} {\gamma'}_{i,j}^{u}
    {\gamma'}_{i+1,j}^{\dagger\, d}{\alpha'}_{i,j+1}^{\dagger\, r} 
    c_{i,j}^k .
    \label{eq:Atilde}
\end{equation}
This way we are able to bring operators containing the same physical fermionic operator together and express the expectation value~(\ref{eq:obs1}) of an arbitrary product operator~(\ref{eq:O}) in a form
\begin{equation}
    \bra{\Psi} O \ket{\Psi} = (-1)^{\sum_{\nu} p_{\nu} P_{\nu} }
    \bra{0}
	K_{m,n}^{[O_{m,n}]} \cdots K_{1,1}^{[O_{1,1}]}
    \ket{0}.
\label{eq:obs2}
\end{equation}
where operators $K_{i,j}^{[O_{i,j}]}$ are obtained by contracting over the physical mode as
\begin{equation}
    K_{i,j}^{[O_{i,j}]} = \ave{{\tilde A'}_{i,j} O_{i,j} A_{i,j} }_{\textrm{phys}}
\label{eq:Kij}
\end{equation}
In the language of tensor networks, this corresponds to obtaining a double-layer structure through contraction over the physical index in two single-layer structure of PEPS. 
Operators $K_{i,j}^{O_{i,j}}$ are of a well defined parity given by the corresponding operator $O_{i,j}$, i.e. $p_{i,j}$.Therefore, the contraction order may be chosen arbitrarily using the fermionic swap rule and anti commutation relations.
In the following we will implicitly assume the dependence of $K_{i,j}^{[O_{i,j}]}$ on the local operator $O_{i,j}$ and use a compact notation $K_{i,j}$.

Due to the canonical anti-commutation relations of fermionic operators, the tensor representation of 
$K_{i,j}$ is not unique. Let us first choose a representation where the norm $\braket{\Psi}{\Psi}$ is 
expressed as a tensor product 
\begin{equation}
\braket{\Psi}{\Psi} = \textrm{tr} \big( \underline{E}^{[1,1]} \cdots \underline{E}^{[1,n]}\cdots \underline{E}^{[m,n]} \big)
\label{eq:normE}
\end{equation}
where the multiplication order is given by the lattice bonds. Such form would enable us to contract the fermionic tensor network exactly.
It is easy to show that this can be achieved by representing operators $K_{i,j}$ defined in~(\ref{eq:Kij}) in the following form 
\begin{equation}
K_{i,j} = E^{[i,j]}_{\ul \,\ur\, \uu\, \ud} \alpha_{i,j+1}^{\dagger\, \ur} \gamma_{i+1, j}^{\dagger\, \ud} \alpha_{i j}^{\ul} \gamma_{i j}^{\uu}
\label{eq:KE}
\end{equation}
where all (local) fermionic signs arising in the process are absorbed in tensor $E^{[i,j]}$.
It should be noted, however, that due to the fermionic signs the matrix
$E_{(l' r' u' d')(l r u d)}$ is no longer positive semi-definite nor Hermitian as is the case in bosonic (spin) systems.
Such an exact contraction scheme is not limited to the calculation of the norm but can be used to contract exactly the expectation value of an arbitrary product operator~(\ref{eq:O}). In such a case, tensors $\underline{E}^{[i,j]}$ in~(\ref{eq:KE}) should be replaced by  ${\tilde E}^{[i,j]}_{\ul\,\ur\,\uu\,\ud}= E^{[i,j]}_{\ul\,\ur\,\uu\,\ud} (-1)^{ \ud \sum_{j'<j} p_{i+1,j'}}$ where $p_{i,j}$ is the parity of $O_{i,j}$ in~(\ref{eq:O}), defined globally.

In order to draw the correspondence with PEPS algorithm, we will choose a different representation of 
$K_{i,j}$ where no signs are produced in contraction over a single row which will allow us to express the boundary row as a matrix product state. This is achieved by the following representation
\begin{equation}
    K_{i,j}^{[O_{i,j}]} = K_{\ul\, \ur\, \uu\, \ud}^{[i,j,O_{i,j}]}
    \alpha_{i,j+1}^{\dagger\, \underline{r} } \gamma_{i+1,j}^{\dagger\, \underline{d}}
    \gamma_{i,j}^{\underline{u}} \alpha_{i,j}^{\underline{l}}
    \label{eq:formK}
\end{equation}
where again all local signs are absorbed in tensor $\underline{K}$. 
For sake of concreteness,  let us write the tensor elements $\underline{K}^{[i,j]}$ explicitly,
\begin{equation}
    K_{\ul\,\ur\,\uu\,\ud}^{[\nu,O_{\nu}]} = 
(-1)^{f_K(\ul,\ur,\uu,\ud)} 
A_{l' r' u' d' k'}^{[\nu]\, *} 
O_{[\nu]}^{k' k}
A_{l r u d k}^{[\nu]}
\end{equation}
with  
$O_{[\nu]}^{k'k} = 
\bra{0}
c_{\nu}^{k'}
O_{\nu}
c_{\nu}^{\dagger\, k}
\ket{0}$ and sign function
$ f_K = l' l + (l' + l)(r + u + d) + (r'+r)(u+d) + d (u'+u) + u' u + u' + l'$.

The double layer structure given by pairs of fermionic operators $\alpha_{i,j}^{l} {\alpha'}_{i,j}^{l'}$ and similar, can also be interpreted as a structure given by higher-dimensional objects $\alpha_{i,j}^{\ul}$ and similar. Evidently, the only property used in the formulation of fermionic network is the parity and all results also apply to higher-dimensional objects $\alpha_{i,j}^{\ul}$ where parity of $\alpha_{i,j}^{\ul}$ is 
given by 
\[
p(\alpha_{i,j}^{\ul}) = \big[ p( {\alpha'}_{i,j}^{l'}) + p( \alpha_{i,j}^l)\big] \textrm{mod} 2.
\]
The fact that the parity is the only relevant element in the anti-commutation relations of operators such as $A_{i,j}$ and $K_{i,j}$, suggests a natural way to generalize the tensor network to higher bond dimensions by replacing virtual fermionic operators $\alpha_{i,j}$ and $\gamma_{i,j}$ in~(\ref{eq:Aij}) by 
higher dimensional objects $\underline{\alpha}_{i,j}$ and $\underline{\gamma}_{i,j}$, respectively.
The rest of the method remains the same whereas all occupation numbers $m$ appearing in fermionic sign factors are replaced by the corresponding \emph{parities} of $\underline{m}$, i.e. 
\(
p(\underline{m}) =  \big(\sum_k m_k\big)\,\mathrm{mod}\,  2 .
\)
The only drawback in such generalization scheme is that it confines the bond dimension to the powers of two. An alternative generalization scheme is by combining fermionic and bosonic (spin) degrees of freedom where the former would assure fermionic nature of description whereas the latter would enlarge the virtual space to capture more entangled physical systems. This would lead to bond dimensions that are even.

\subsection{Efficient contraction of fermionic tensor network}
Fermionic PEPS can be exactly contracted in a sign-free way using a suitable representation of operators $K_{i j}$ as shown in~(\ref{eq:normE}). 
However, exact contraction is only possible with small bond dimension and small lattice sizes since the complexity scales exponentially with the linear lattice dimension and we have to resort to an approximate contraction scheme \cite{peps}.
Let us quickly review the approximation scheme to calculate expectation values as used in the PEPS algorithm. 
The first and the last rows are recognized as matrix product states $\xi_1$ and $\xi_n$, respectively, 
and all inner rows correspond to matrix product operators $\Xi_j$ for ${j\in\{2,3,\ldots,n-1\}}$.
The expectation value $\bra{\xi_n} \Xi_{n-1} \Xi_{n-2} \cdots \Xi_2 \ket{\xi_1}$ is 
calculated by approximating a product $\Xi_2 \ket{\xi_1}$ by a new matrix product state $\ket{\xi_2}$ of some finite bond dimension \cite{verstraetempo} and proceeding iteratively until the expectation value is
given by $\bra{\xi_{j+1}} \Xi_j \ket{\xi_{j-1}}$.

In the following we will show how the expectation value of an arbitrary product operator can be calculated efficiently in an approximate way, which is equivalent to the approach in occupation number representation \cite{eisert2} or tensor network approach \cite{corboz2}. 
Taking the advantage of representation~(\ref{eq:formK}) of $K_{i j}$ where no signs are produced in the horizontal contraction we express the first row as a matrix product state \begin{equation}
\ket{\xi_1} = 
{\rm tr} \big( \underline{K}^{[1,1] \ud_1} \cdots \underline{K}^{[1,n] \ud_n} \big)
\gamma_{2,n}^{\dagger\, \ud_n}\cdots \gamma_{2,1}^{\dagger\, \ud_1}\ket{0}
\label{eq:MPS}
\end{equation}
with matrices $(\underline{K}^{[1,j] \ud} )_{\ul, \ur} = K_{\ul,\ur,0,\ud}^{[1,j,O_{1,j}]}$. The same applies to the last row.
Inner rows, on the other hand, cannot be represented as matrix product operators in a form which would allow immediate contraction with matrix product states due to the fermionic signs produced by reordering vertical virtual fermionic operators. 
\begin{figure}
\centering
\includegraphics[width=0.8\columnwidth]{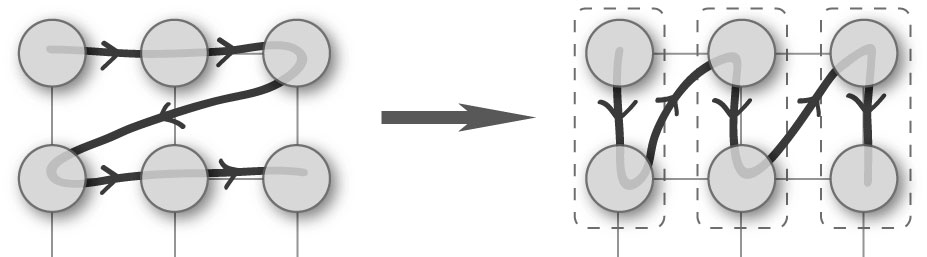}
\caption{Merging two rows together (\ref{eq:merge2rows}) and replacing the double row by a MPS.}
\label{fig:merge2rows}
\end{figure}
Nevertheless, using the fact that the parity of $K_{i,j}$ is determined globally
by the underlying operator $O_{i,j}$, 
one can change the contraction order in contracting first two rows to 
\begin{equation}
K_{2 n} \!\cdots\! K_{2 1}K_{1 n} \!\cdots\! K_{1 1} %
= 
(-1)^{f} K_{ 2 n}K_{1 n}\! \cdots \!K_{2 1}K_{1 1}
\label{eq:merge2rows}
\end{equation}
where $f = \sum_{i=1}^n p_{1 i}\sum_{j=1}^{i-1}  p_{2 j}$. Note that this step is trivial since there is no need for fermionic swap rules as no fermionic modes are crossed.
Contracting products $K_{2 j} K_{1 j}$ over the vertical mode (Fig.~\ref{fig:merge2rows}) and representing the result in form~(\ref{eq:formK}), we again obtain a matrix product description of form~(\ref{eq:MPS}). 
This way, the fermionic nature is completely absorbed in the tensors and all the MPS formalism results apply.

\subsection{Variational simulation of the ground state}

There are essentially two ways of simulating the ground state using tensor networks. 
The first possibility is the evolution of a PEPS state in imaginary time using the approximate Trotter decomposition of the evolution operator.
The alternative way is to optimize tensors $\underline{A}^{[i,j]}$ site by site in a variational way such that the total energy
\(E = \bra{\Psi} H \ket{\Psi} / { \braket{\Psi}{\Psi} } \)
is minimal.
While numerical stability often speaks in favor of the imaginary time evolution, the variational approach is faster and gives fairly good results after a single optimization sweep over the lattice. In this paper, we shall only focus on the latter approach and show that all fermionic signs which appear in the computation are absorbed locally into tensors which makes the problem essentially sign-free for practical matters and thus well suited to conventional PEPS techniques.

In the following we will show how to write the total energy as a function of a tensor $\underline{A}^{[i,j]}$ in a sign-free way.
Using the fermionic rule it is easy to show that the expectation value ~(\ref{eq:obs2}) of an arbitrary product operator can be written as
\begin{equation}
\bra{\Psi} O \ket{\Psi} = \bra{0} {\breve A}_{i,j}^\dagger \Omega_{i,j}^{[O]} {\breve A}_{i,j} \ket{0}
\label{eq:omega}
\end{equation}
where ${\breve A}_{i,j}$ contains all tensor elements of $\underline{A}^{[i,j]}$ as
\[
{\breve A}_{i,j} = A_{l,r,u,d,k}^{[i,j]} c_{i,j}^{\dagger \, k} \alpha_{i,j}^{\dagger\, l} \alpha_{i,j+1}^{\dagger\, r} \gamma_{i,j}^{\dagger \, u}\gamma_{i+1,j}^{\dagger\, d}
\]
 whereas $\Omega_{i,j}^{[O]}$ contains tensors corresponding to all other lattice sites. Such expression may be easily obtained by replacing $K_{i,j}$ in~(\ref{eq:obs2}) according to~(\ref{eq:Kij}) and anti-commuting ${\tilde A}_{i,j}$ and $A_{i,j}$ to the far ends. Finally, all fermionic signs are absorbed in $\Omega_{i,j}^{[O]}$.
Note that the sign in~(\ref{eq:obs2}) is cancelled by commuting ${\tilde A'}_{i,j}$ defined in 
(\ref{eq:Atilde}) over all consequent sites and thus no longer appears in~(\ref{eq:omega}). 
Using a convenient representation for $\Omega_{i,j}^{[O]}$, i.e.
\begin{eqnarray}
    \Omega_{i,j}^{[O]} &=& \Omega_{l' r' u' d' k' l r u d k}^{[O,i,j]} \times \label{eq:omegaform} \\ 
    &\times& {c'}_{i,j}^{\dagger\, k'} {\alpha '}_{i,j}^{\dagger\, l'} {\alpha '}_{i,j+1}^{\dagger\, r'} {\gamma '}_{i,j}^{\dagger\, u'} {\gamma '}_{i+1,j}^{\dagger\, d'}  
\gamma_{i+1,j}^d \gamma_{i,j}^u \alpha_{i,j+1}^r \alpha_{i,j}^l c_{i,j}^k \nonumber
\end{eqnarray}
we are able to rewrite the expectation value as an ordinary scalar product
\begin{equation}
    \bra{\Psi} O \ket{\Psi} = \underline{A} \cdot \underline{\Omega} \, \underline{A}
\end{equation}
where vector elements of $\underline{A}$ are given by 
$A_{(l r u d k)}^{[i,j]}$ and similarly for matrix elements of $\underline{\Omega}$ given as $\Omega_{(l' r' u' d' k')(l r u d k)}$. 
This way the expectation value of a fermionic operator is expressed in terms of a sign-free linear algebra expression.
Note however that the initial assumption that tensors $A_{l r u d k}^{[i,j]}$ are of 
well defined parity, reduces the effective subspace of the vector space to the
even-even or odd-odd sector with respect to indices $(l' r' u' d' k')$ and $(l r u d k)$.
While operator $\Omega$ itself is always of even parity, 
i.e. $p(l' r' u' d' k' l r u d k) = 0$, no such requirement is imposed separately
to $(l' r' u' d' k')$ and $(l r u d k)$. 
In principle, both sub-sectors, even-even and odd-odd, should
be obtained separately using the assumption for the parity of ${\breve A}$ or equivalently, tensor elements $A_{l r u d k}^{[i,j]}$. 
However, since $O$ is in total of even parity, no additional signs are produced in the
odd-odd case where $\Omega_{i j}^{[O]}$ is represented in form~(\ref{eq:omegaform}). 

The total energy $\braket{\Psi} H \ket{\Psi}/\braket{\Psi}{\Psi}$ may be expressed in terms of effective operators as
\begin{equation}
E = \frac{ \underline{A} \, \cdot \underline{H}_{\rm eff} \, \underline{A} } %
{\underline{A} \cdot \underline{N}_{\rm eff}  \, \underline{A} }
\label{eq:totalenergy}
\end{equation}
where $\underline{N}_{\rm eff}$ and $\underline{H}_{\rm eff}$ are obtained using the above described procedure for the identity operator and the Hamiltonian operator, respectively, where the latter is written as a superposition of product operators. Note that the computation of $\underline{H}_{\rm eff}$ is simplified for Hamiltonians with local interactions where certain operators are grouped together in the (approximate) contraction process.

The solution $\underline{A}$ which minimizes~(\ref{eq:totalenergy}) is formally given by the lowest eigenvalue solution of a generalized eigenvalue problem 
\begin{equation}
\underline{H}_{\rm eff} \, \underline{A} = \lambda \underline{N}_{\rm eff}\,  \underline{A}
\label{eq:geneigproblem}
\end{equation}
where $N_{\rm eff}$ is a semi-definite Hermitian matrix and $H_{\rm eff}$ is Hermitian.
Due to the parity constraints, the eigenvalue problem must be solved separately for both parity sub-sectors and the better solution should be retained.
The generalized eigenvalue problem~(\ref{eq:geneigproblem}) is only well defined if $N_{\rm eff}$ is nonsingular. In one-dimensional variational MPS with open boundary conditions one can always renormalize the tensor network in a way that $N_{\rm eff}$ is exactly equal to the identity which simplifies the computation and, more importantly, makes the method stable.
In two dimensions, the way to make PEPS better conditioned remains an open question.
In general, the spectrum of $N_{\rm eff}$ might and does contain very small values or even zeros, in which case the standard algorithm would produce infinite or ill-disposed eigenvalues. 
The ill-conditioned generalized eigenvalue problem must be solved in an approximate fashion by isolating such invalid solutions either by projecting out the null-space of $\underline{N}_{\rm eff}$ 
or using more sophisticated algorithms such as Fix-Heiberger reduction \cite{fixheiberger} where ill-conditioned modes of $\underline{N}_{\rm eff}$ are not completely neglected.

We find that the most stable way is to project the system to the subspace spanned by well conditioned eigenvectors of $N_{\rm eff}$ with respect to a cutoff $\delta$ and then using the Fix-Heiberger algorithm with the tolerance $\epsilon \gtrsim 10 \delta$ which eliminates all solutions unstable to the perturbation of $\epsilon$ to the matrices $\underline{H}_{\rm eff}$ and $\underline{N}_{\rm eff}$. In addition, when a good convergence is achieved, we optimize the total energy~(\ref{eq:totalenergy}) in an iterative way using the conjugate-gradient method.
Nevertheless, compromise between efficiency and accuracy versus numerical stability must be made.

\section{Results}
\begin{figure}
    \centering
    \includegraphics[width=0.98\columnwidth]{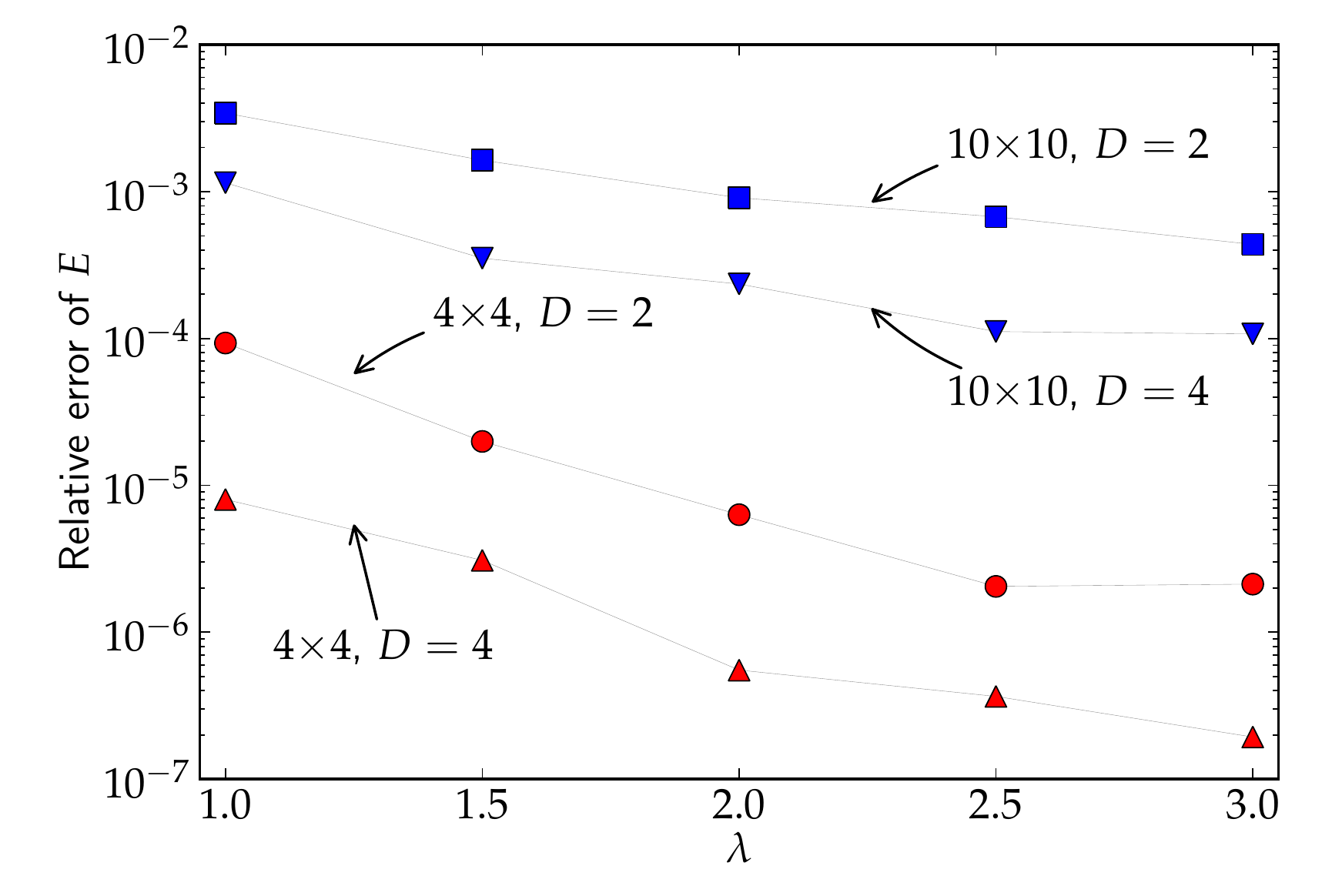}
    \caption{Relative error of the ground state energy for the quadratic model~[Eq. \ref({eq:H1})] with $\gamma=1$ for lattice sizes $10{\times}10$ and $4{\times}4$ and bond dimension $D=2$ and $D=4$. The truncation number is in all cases set to ${\tilde D}=64$.}
    \label{fig:1}
\end{figure}

The finite size fermionic PEPS method is put to the test by simulating an exactly solvable bi-linear (quadratic) model on a square lattice. The model consists of three parts: hopping between nearest neighbor, pair creation/annihilation and chemical potential, described by the following Hamiltonian operator \cite{liprb}
\begin{equation}
    H = \sum_{\langle \mu\nu\rangle}
        \big[ c_{\mu}^\dagger (c_\nu - \gamma c_\nu^\dagger) + \textrm{h.c.} \big] 
    - 2 \sum_{\nu} \lambda c_{\nu}^\dagger c_{\nu} .
    \label{eq:H1}
\end{equation}
The pairing potential $\gamma \geq 0$ is used to destroy the total particle number symmetry and $\lambda \geq 0$ is the chemical potential.
The same model was also used in \cite{corboz2} where infinite fermionic PEPS algorithm was presented. Unlike Refs.~\cite{liprb,corboz2} we assume open boundary conditions which is better suited for finite-size PEPS algorithm.
The system is critical for $\lambda \leq 2$ (gapless in the thermodynamic limit) and non-critical (gapped) elsewhere. We choose a line $\gamma=1$, $\lambda \in [1, 3]$ in the parameter space and test the method with respect to the relative accuracy of the ground state energy as shown in Fig.~\ref{fig:1}. 
For the bond dimension we take either $D=2$ or $D=4$ which corresponds to one or two virtual fermions of each kind, respectively. The maximal bond dimension in the process of contracting the double-layer structure (see Ref.~\cite{valentin} for details) is designated by the truncation number ${\tilde D}=64$ which will be justified later. As expected, the method performs better in the gapped regime ($\lambda > 2$) for both system sizes considered in Fig.~\ref{fig:1}. 
For the $10\times 10$ lattice the spectral gap in the gapped regime at $\gamma=1$, $\lambda=3$ is of order of $3 \cdot  10^{-3} \vert  E_0\vert$ where $E_0$ denotes the corresponding ground state energy and the total energy obtained from the simulation is below the energy of the first excited state. In the gapless regime, e.g. for $\gamma=\lambda=1$, no guarantee for the ground state is given since the spectral gap is of order of $10^{-7} \vert E_0\vert $. For the $4 \times 4$ lattice the spectral gap in the gapless regime is of order $10^{-3} \vert E_0 \vert$ which is a magnitude larger than the accuracy of the ground state energy.

By increasing the bond dimension from $D=2$ to $D=4$, the relative accuracy is improved for an order of magnitude as seen in Fig.~\ref{fig:1} for both lattice sizes $4{\times}4$ and $10{\times}10$. 
Note however, that a fairly good precision is achieved already with the bond dimension $D=2$. 
The algorithm would perform better for higher bond dimension if one could make PEPS well conditioned.
Namely, with the increasing bond dimension the problem~(\ref{eq:totalenergy}) becomes less conditioned and it is essential to use Fix-Heiberger procedure (and conjugate gradient method) to eliminate unstable solutions. 
If all nearly singular vectors were simply chopped away, the benefit of using higher bond dimension would be negligible.

\begin{figure}
\includegraphics[width=0.98\columnwidth]{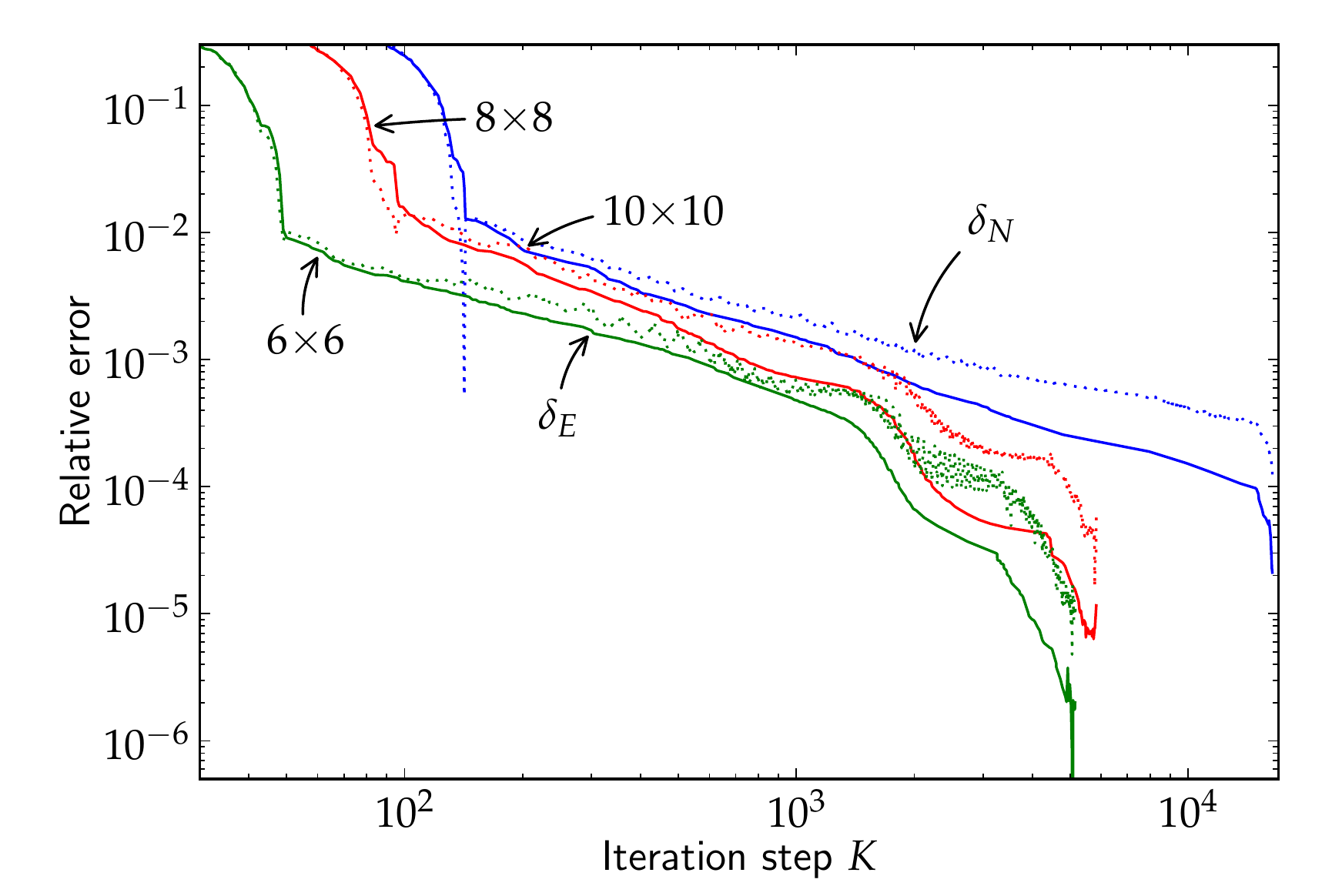}
\caption{Convergence in terms of the ground state energy relative error $\delta_E$ (full lines) and the total particle number relative error $\delta_N$ (dotted lines) for (\ref{eq:H1}) with $\gamma=1$ and $\lambda=3$. Three lattices sizes are considered as designated in the legend. Bond dimension is taken $D=4$ with the truncation number ${\tilde D}=64$.}
\label{fig:2}
\end{figure}
In Fig.~\ref{fig:2} we show the convergence of the ground state energy and the total particle number as a function of the number $K$ of single particle optimizations. 
The simulation is done first using the bond dimension $D=2$ and switching to $D=4$ when sufficiently good convergence rate (relative difference $10^{-5}$ for the total energy between two consequent \emph{sweeps}) is achieved.
We consider three lattices sizes and observe that a fairly good approximation to the ground state where the ground state energy is accurate to $1\%$, is achieved with less than two sweeps over the lattice. The initial state was in all cases taken random. 
\begin{figure}
\includegraphics[width=0.98\columnwidth]{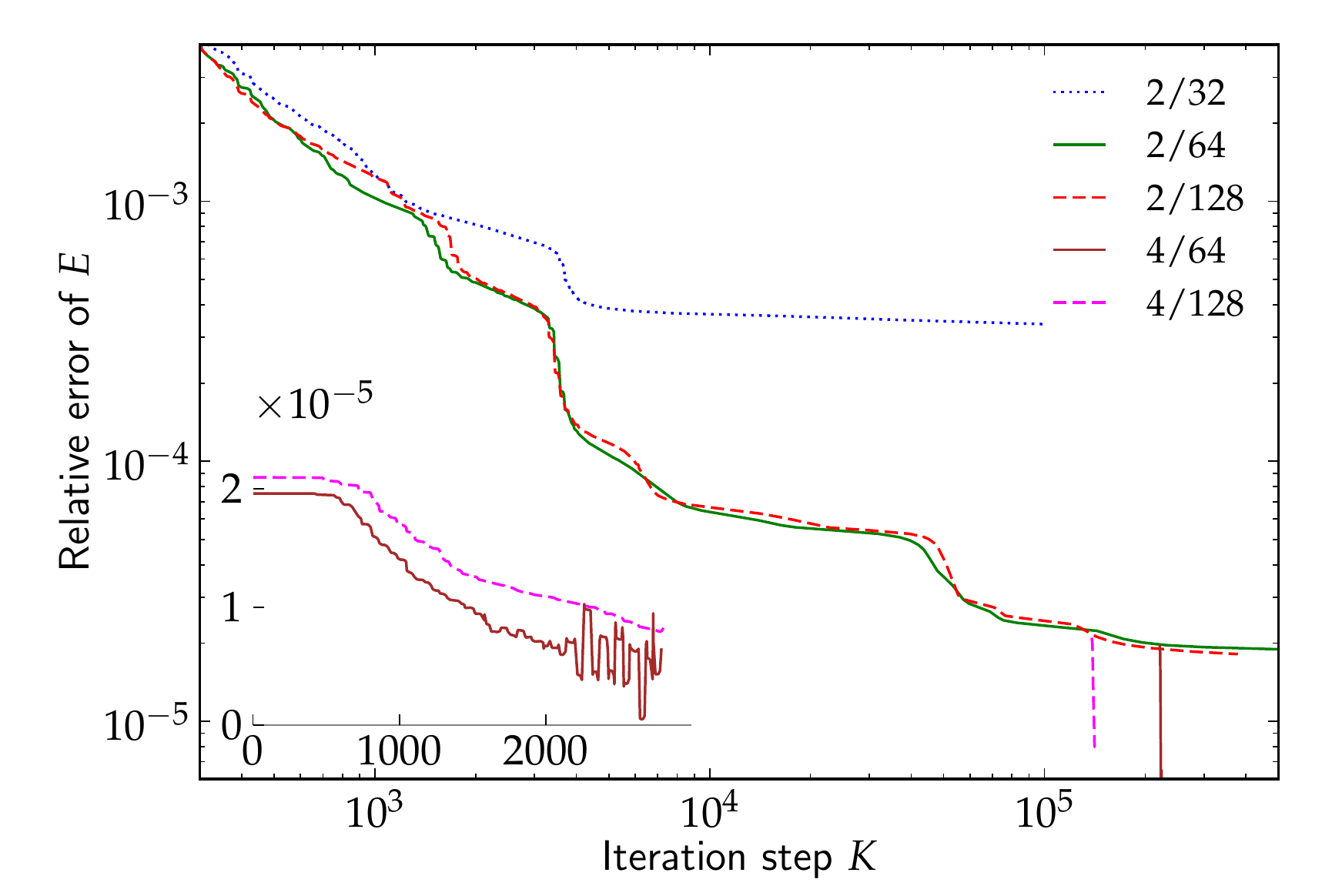}
\caption{Convergence for the ground state energy for~(\ref{eq:H1}) with $\gamma=1$, $\lambda=2.5$ on a $8{\times}8$ lattice. Legend entries designate $D/{\tilde D}$. Results for $D=4$ are obtained starting from a good approximation for $D=2$ (almost vertical lines on the main plot, magnified on the left in the linear-linear scale).
}
\label{fig:5}
\end{figure}
After the initial sweep the convergence becomes slower but the relative error of both the ground state energy and the total particle number typically decays as $1/K$. In variational methods such as PEPS the ground state energy is typically more accurate than other observables such as the total particle number which is also confirmed in Fig.~\ref{fig:2}. 
We have however no explanation for the oscillations in accuracy for the total number of particles.

Let us now check the validity of the results for various truncation numbers ${\tilde D}$ used to truncate the large matrix products representing several consecutive rows. 
As presented in Fig.~\ref{fig:1}, we used ${\tilde D}=64$ which turned out to be sufficient to get good accuracy. 
In Fig.~\ref{fig:5} we present the results for the quadratic model a $8{\times}8$ lattice where the same initial state was taken in all cases. 
We consider three different values of ${\tilde D}$ for the bond dimension $D=2$.
Eventually, we start the simulation with the bond dimension $D=4$ where the (almost converged) results from $D=2$ were taken as the initial state, also magnified on the left side of Fig.~\ref{fig:5}.
We observe that ${\tilde D}=32$ is insufficient to achieve good accuracy of the ground state energy although it gives reasonable results with little effort.
There is virtually no difference between the cases ${\tilde D}=64$ and ${\tilde D}=128$ except the latter being computationally much more demanding.
As already mentioned in the previous section, the algorithm eventually produces unstable solutions where the effective norm operator $\underline{N}_{\rm eff}$ in~(\ref{eq:geneigproblem}) becomes more and more ill-conditioned.
This is reflected in the oscillations seen in the magnification of Fig.~\ref{fig:5} which are also a sign that the simulation should be stopped, unless the state is made better conditioned.

\section{Discussion}
The finite size fermionic PEPS method was tested on a trivial example of a quadratic integrable model where it was shown that fairly good results can be achieved for lattice sizes $10{\times}10$. The present formulation is however open to various improvements and modifications.
The first improvement would be beneficial not just for fermionic PEPS but for all two dimensional PEPS Ans\"{a}tze, namely a way to make PEPS better conditioned which is of crucial importance for employing higher bond dimensions. Another possibility would be to consider higher order symmetries such as the $Z_k$ symmetry for which the presented $Z_2$ symmetry algorithm presents a good starting point.

The fermionic swapping rule allows arbitrary manipulations to the contraction order which enables various enhancements to the presented method. The first is a complementary way of optimizing tensors $\underline{A}^{[i,j]}$ by imaginary time evolution.
The method can also be made more robust in convergence to the global minimum by adding stochastic updates to the tensor elements which would be beneficial especially with non-trivial models where the energy landscape is such that one easily gets stuck in a local minimum. Although such phenomenon was not observed in simulating the integrable model~(\ref{eq:H1}), it might occur for certain interacting models.
Let us briefly consider an interacting model
\begin{equation}
H = -\sum_{\langle \nu \mu\rangle } \big[
c_{\nu}^\dagger c_{\mu}  + {\rm h.c.}
\big]
+ 
V \sum_{\langle \nu \mu\rangle } n_\nu n_{\mu}
\label{eq:Hint}
\end{equation}
where the total particle number $\langle N \rangle$  for 
$N = \sum_{i,j} c_{i,j}^\dagger c_{i,j}$ is a preserved quantity. The algorithm does not always converge to the global ground state but to the lowest-lying eigenstate in a particular total-particle number sub-sector, depending on the initial state. This issue may be addressed by simulating a modified model $H' = H - \mu N$ with the same eigenstates as~(\ref{eq:Hint}). Various total particle number sub-sectors are achieved by tuning the chemical potential.

\begin{figure}
\includegraphics[width=0.98\columnwidth]{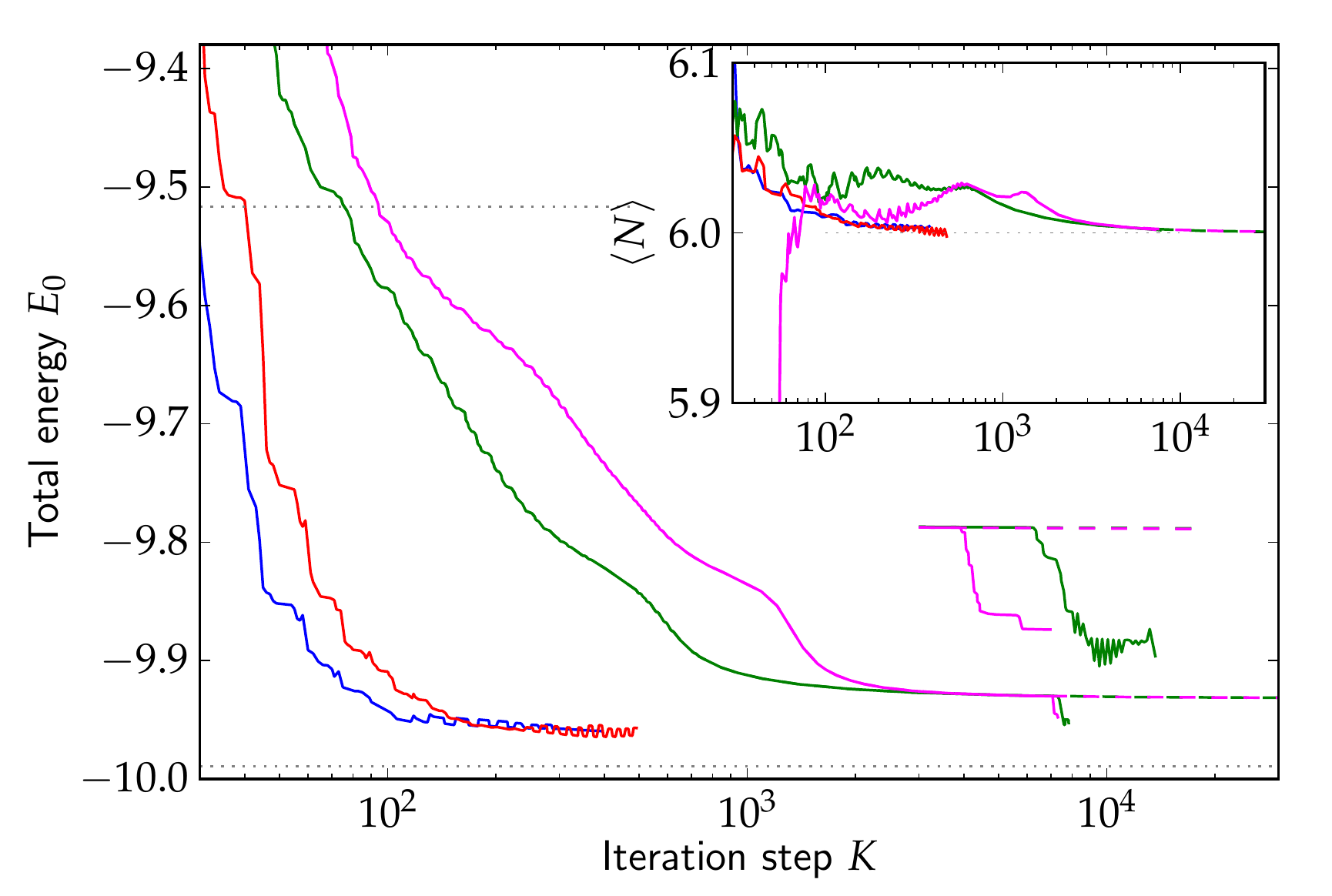}
\caption{(Color online.) Convergence of the total energy $E_0$ for the interacting model~(\ref{eq:Hint}) with $V=0.5$ on a $4{\times}4$ lattice. Initial states are random with $D=4$ (first two curves from the left), $D=2$ (two curves on the right), and $D=2$ switched to $D=4$ (small deviation from the $D=2$). The transition $D=2$ to $D=4$ is magnified in the plot. The inset shows the total particle number $\langle N\rangle$ for the corresponding cases. The truncation number is set to ${\tilde D}=128$ in all cases.}
\label{fig:3}
\end{figure}
In Fig.~\ref{fig:3} we present the total energy convergence for the interacting model~(\ref{eq:Hint}) on a $4{\times}4$ lattice with the interaction strength $V=0.5$.
Different lines correspond to different random initial states and bond dimensions as explained in the figure caption.
We observe that in this case a quick convergence to the global minimum (note the exact energy levels designated by dotted lines) is achieved for all choices of random initial state although the relative precision is not as good as in the integrable model.
A higher bond dimension $D=4$ gives better results after fewer number of iteration steps but effectively consumes more computational time.
Similar accuracy is obtained when an approximate ground state is obtained by a small bond dimension $D=2$ and later switched to $D=4$.
The simulation however quickly stops due to achieved relative accuracy between subsequent sweeps.
The fluctuations in the energy in $D=4$ are explained by the transitions between rows when an error is made in truncating large matrix product states (note that the energy is calculated in an approximate way).
The total particle number is in all cases in agreement with the exact value in the ground state, $\langle N \rangle = 6$ for this choice of parameters.
It must be noted that not every initial state converges to the ground state but might as well converge to a local minimum.
No such case was however observed for $V=0.5$ on a $4{\times}4$ lattice.
It might be beneficial to tune the chemical potential to influence the number of particles in the system or start with a good initial state.

\begin{figure}
\includegraphics[width=0.98\columnwidth]{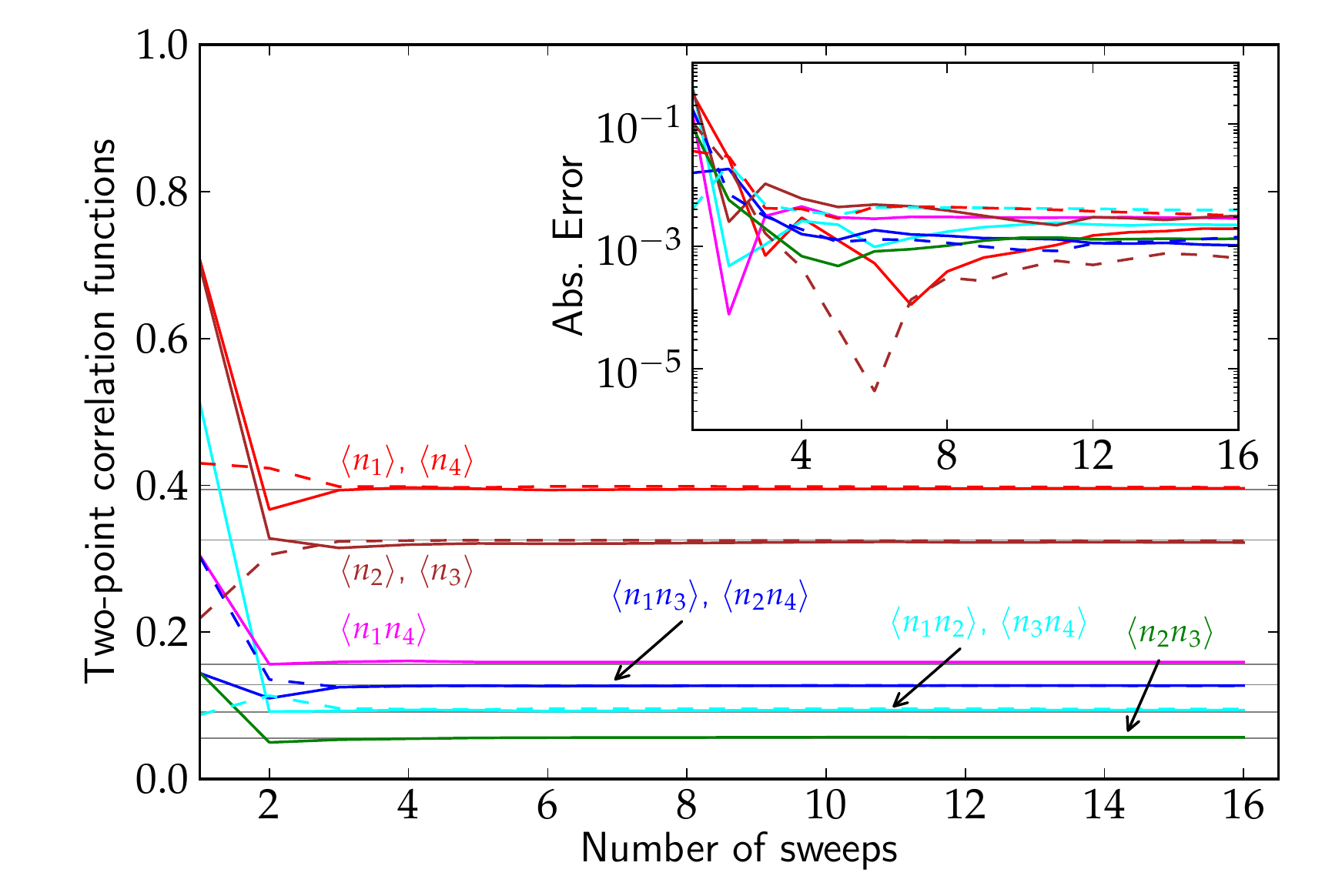}
\caption{(Color online.) Correlation functions $\langle n_{3,j} n_{3,j'} \rangle$ (designated as $\langle n_j n_{j'}\rangle$ in the graph, the second label is shown by dashed lines) for the interacting model~(\ref{eq:Hint}) with $V=0.5$ on a $4{\times}4$ lattice. The data correspond to the left-most curve in the graph on Fig.~\ref{fig:3}. Exact values are plotted by thin gray lines.}
\label{fig:4}
\end{figure}
Apart from observables consisting of local contributions such as the energy or the total particle number, we can also investigate nonlocal quantities such as correlation functions, e.g. the density-density correlations $\langle n_{i,j} n_{i',j'}\rangle$.
In Fig.~\ref{fig:4} we show the density-density correlations for a fixed row ($i=3$) for the interacting model~(\ref{eq:Hint}) on a $4{\times}4$ lattice.
The data correspond to the first curve from the left in Fig.~\ref{fig:3}.
For computational simplicity we only calculate the correlations after each complete sweep, i.e. every $(2 m n -2)$th step.
It can be seen that in all cases the results practically coincide with the exact results designated by thin gray horizontal lines with the absolute error of order of $10^{-3}$ as shown in the inset.

\section{Conclusion}
We have presented a finite size fermionic PEPS method to simulate ground states of two dimensional fermionic systems \cite{kraus} completely in terms of fermionic operators.
Using a fermionic swap rule to reverse the contraction order of two superpositions of products of fermionic canonical operators we have shown how a fermionic tensor network can be contracted exactly without introducing any additional sign bond but instead absorbing all signs locally.
Due to the parity constraints in fermionic systems we have presented a way, equivalent to \cite{eisert2,corboz2}, to calculate the expectation values for arbitrary operators efficiently in an approximate fashion.
Finally, we have implemented the variational PEPS algorithm on a fermionic lattice and tested it on an integrable bi-linear fermionic model.
We have found that the ground states of such a model can be simulated efficiently with relatively high accuracy in the ground state energy and the total number of particles.
We have also discussed the performance of the method in the case of an interacting fermionic system where the method converges to the global minimum, albeit with less accurate precision.
Besides local observables such as the energy and the total number of particles, the method correctly describes also the non-local two-point correlations.

\begin{acknowledgments}
I.P. acknowledges fruitful discussions with V. Murg. This work was supported by the FWF grant FoQuS and the ERC grant QUERG.
\end{acknowledgments}

\end{document}